# A STUDY OF THE INTERFACE USABILITY ISSUES OF MOBILE LEARNING APPLICATIONS FOR SMART PHONES FROM THE USER'S PERSPECTIVE

Abdalha Ali, Muasaad Alrasheedi, Abdelkader Ouda and Luiz Fernando Capretz

Department of Electrical and Computer Engineering, Western University, London Ontario, N6A5B9, Canada

## ABSTRACT

 A conceptual framework for measuring the usability characteristics of mobile learning (m-Learning) application has been developed. Furthermore, a software prototype for smartphones to assess usability issues of m-Learning applications has also been designed and implemented. This prototype has been developed, using Java language and the Android Software Development Kit, based on the recommended guidelines of the proposed conceptual framework. The usability of the proposed model was compared to a generally available similar mobile application (based on the Blackboard) by conducting a questionnaire-based survey at Western University. The two models were evaluated in terms of ease of use, user satisfaction, attractiveness, and learnability. The results of the questionnaire showed that the participants considered the user interface based on our proposed framework more user-friendly as compared to the Blackboard-based user interface.

## 1.INTRODUCTION

Mobile devices are increasingly becoming integrated into various aspects of our daily lives. One area is in the educational sector, where mobile phones are being used as the platform for teaching and learning. However, unlike personal computers, the screen size and resolution restrict mobile phones in displaying content [1]. Learning by using specifically smartphones, is being integrated within existing education systems to support real-time communication and deliver learning materials.

For instance, smartphones are being used in many universities as a classroom tool to engage and support students in communicative, collaborative, supportive, and constructive activities. Additionally, mobile technologies enable individual learners to build knowledge and construct understandings; in this they facilitate a change in the pattern of work activity/learning [2].

However, mobile applications used for educational purposes have a complex user interface (UI) with many hidden options. There is already a great interest in designing and developing attractive, user-friendly mobile applications to gain the acceptance of end user. Further, in order to be acceptable to a wider audience, the applications need to be both robust and of a very high quality [3].

Due to the significant diffusion of mobile technologies, most students today already own mobile devices. Hence, the technology is a strong contender to be the next "big thing" in educational platforms [4]. Mobile technology can deliver educational content in several ways. For instance, Wang et al. [5] reported that mobile phones could be used to deliver online courses to university students. In fact, the multitude of ways in which mobile technology can be used in the educational sector, prompted Prensky [6] to note that students will be able to learn "anything, if developers designed it right".

Also, the demand for learning anywhere and anytime has specified the need for a new type of electronic learning known as m-Learning to take advantage of mobile devices which are becoming increasingly popular [7]. M-learning is an education mode in which students can use mobile communication terminals to assist them in learning [8].

One of the key components of a successful and acceptable educational application is ease of use. Several high quality applications available in the market actually lose out because of their complex application and unattractive and confusing user interface. Thus, when designing a user interface for mobile phones, especially for education purposes, the user requirements of these devices should be considered [3].

Many usability guidelines are used for designing desktop applications [9][10]. However, these guidelines cannot be utilized to design and develop m-learning applications, simply because neither addresses the aspect of mobility not the obvious limitations of the mobile devices, like such as screen size, and the need for wireless connectivity [1]. There is a singular lack of reliable usability guidelines, specifically meant for designing and developing m-learning with user-friendly interfaces. While research conducted on the success factors of m-Learning clearly show that usability and related aspects are one the core requirements, the specific ways in which this can be addressed is are lacking [12]. In fact, usability has been less extensively covered than the technological aspects of the m-learning. Mobile technology can be successful as an educational platform only when the future research into the area of m-Learning includes fruitful discussion in of all the aspects of usability: – learnability, understandability, ease of use, effectiveness, and efficiency of mobile applications [13].

The experience of end users is extremely important to the success of m-Learning. Therefore, a mobile phone application targeted towards education must be designed and developed keeping in mind ease of use, usefulness, and attitude and intention to use that will help ensure a high level of acceptance [14]. Assessment must be conducted to precisely identify where improvement is required. Therefore, the main purpose of this research study is to develop a framework for measuring the quality aspects of m-Learning. Using this framework, existing m-Learning applications can be assessed in terms of their usability. In addition, a prototype application has been developed composed of two user interfaces, Model A and Model B. Model A has been designed and developed based on a user interface adopted from the Blackboard website [15], while Model B has been developed following the Android SDK recommendation and using the proposed framework as a guideline. The prototype has been used for assessing the framework by empirical evaluation, validation, and comparison of usability issues of the model using the proposed framework as a guideline.

The paper is organized into the following sections. Section 2 presents the literature review on the subject as well as work in related areas. Section 3 describes the development and theoretical assessment of the theoretical framework. Section 4 presents the assessment methodology used for evaluating the framework. Section 5 presents the results of an analysis, and the discussion of the results is presented in section 6. Research conclusions and possible directions of future research work are presented in section 7.

## 2.LITERATURE REVIEW

In this section we discuss literature that deals with the usability and summarize a selection of the most relevant findings. To start, in the ISO 9241-11 (1997) [16] standard, usability is defined as "the extent to which a product can be used by specified users to achieve specified goals with effectiveness, efficiency and satisfaction in a specified context of use". However, ISO/IEC 9126-1 (2001) [17], states that usability is "the capability of the software product to be understood, learned, used and attractive to the user, when used under specified conditions." [18], on the other hand, emphasize that there is a great deal of literature available that addresses usability, user interface design, and related topics for mobile devices. A mobile application must be developed and designed with respect to user technological ability, skills, and language proficiency. This forces developers to be very careful with design issues in order to maximize the level of usability with all of its sub-characteristics. Ziefle and Bay [19] demonstrate that awareness of user interface structure is one of the most important issues concerning cell phones.

On the other hand, Jarvela et al. [20] studied how to help users participate in collaborative learning using smartphones. The researchers utilized a mobile lecture interaction tool to encourage students in higher education to participate in a class discussion. This tool enabled participants to ask and answer questions, as well as to rate classmate questions. The main purpose of this survey was to get student feedback on the usability of the tool. The feedback showed that mobile tools with a high level of usability will definitely increase their engagement in discussions. Mobile technology allows the users to communicate instantly; this characteristic plays a vital role in a successful m-Learning environment. However, usability issues are found to be important factors in the learners' high satisfaction level with the cooperative learning available through the system.

The Mobile System Analysis and Design (MOSAD) application [21] is a mobile application used as a revision tool for the System Analysis and Design (SAD) course at University Technology Petronas. The researchers' main goal was to design an m-Learning application that allows students to review and read notes during their spare time, and more importantly, to evaluate this application by considering some design issues that could be modified to improve its usability. After the application was designed, a heuristic evaluation was completed to measure its level of usability. Many tests were conducted, and the purpose of those tests was to receive feedback from participants so the level of usability of this application could be determined. The results indicate that adding some features to the design will be useful and will improve the overall usability of the application.

## 3.DEVELOPMENT OF THEORETICAL FRAMEWORK

In this section, the development of framework for assessment of the m-Learning platform is described. First, the technical and non-technical quality aspects, the basis of an m-Learning platform assessment, are discussed. This information will be used to develop the actual framework. A general theoretical analysis of the framework is also presented in this section.

### 3.1.Development of Theoretical Framework Assessing Usability of M-Learning Platform

The framework is shown in Figure 1 below. Basically, this framework is a combination of structural factors [23]: rules, goals, outcomes, competition, interaction, and representation. It also integrates dimensions of the learning context [24]: identity, time-location, facility (mobile phones), activity, learner on the move, and community. A similar framework has been introduced by Parsons and Ryu, [13]; however, we have identified three design issues: usability, communication, and interactivity.

These three design issues can be added to previous design issues identified by Parsons and Ryu [13]: learning on the move, media type, collaboration, user role, and profile. First, these design issues have been linked to the dimensions of learning context and then to the structural factors. From these two steps we target social skills and team building, new knowledge, and improved skills. Key features of the framework are that it identifies the importance of design issues, the dimensions of the learning context, and structural factors in order to address learning objectives that have both a user focus and a platform focus. The usability design issue, which includes ease of use and understanding, can be achieved if users are able to use the applications (usability design issue). This can also be achieved by using identifiers of each user that must be unique in the name space, and that are accessed by the application via a log-on system (with a user name, password, and/or special devices such as smart cards or fingerprint reader). All these identifiers are classified under the "identity dimension" within the learning context. Furthermore, ease of use and the ability to utilize specific identifiers for each user will enable them to perform tasks – such as reviewing the lessons, doing assignments, and participating in group sessions with other users – with, of course, the support of mobile devices (structural factors: business rules and learning roles).

Finally, the learning objectives are addressed; these may include improving skills, acquiring new knowledge, social skills, or team building. However, as we have mentioned in this paper, each design issue may be linked with more than one of the dimensions within the learning content. In summary, most of the components of this framework will relate to each other in one way or another. For example, usability factors can be affected by more than one of the learning contexts that include identity, learner, activity, time-location, facility (mobile phone), and the software engineering community [25] [26]. On the other hand, each dimension of the learning context may be linked to more than one of the structural factors, which in turn, are linked to different learning objectives.

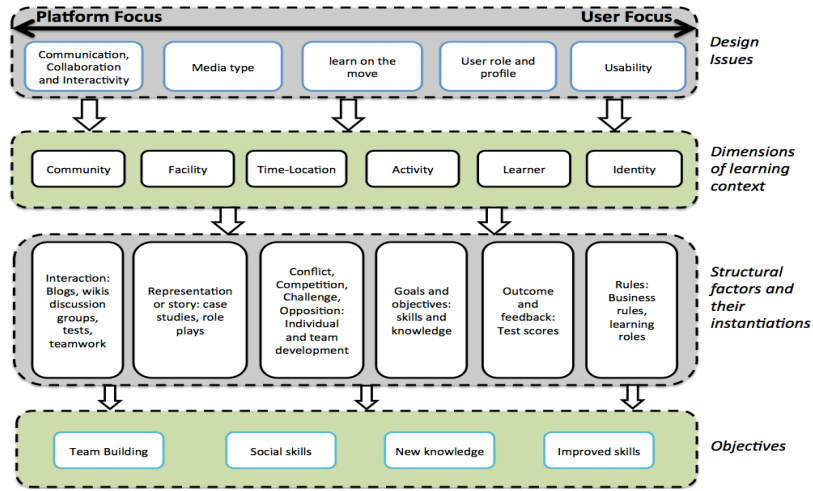

Figure 1. A Conceptual Framework for Measuring Quality Aspects of m-Learning.

## 3.2. Theoretical Analysis of Proposed Framework

The relationships among the quality of use, internal quality, and external quality metrics are shown in Figure 2 below. We can see that if extensions are applied to the ISO/IEC metrics [18], it is possible to map these metrics with our framework in order to measure the design issues related to learning on the move, user roles and profiles, media type, and usability issues. However, additional metrics are needed to complement the context of the use dimensions of the quality in use metrics.

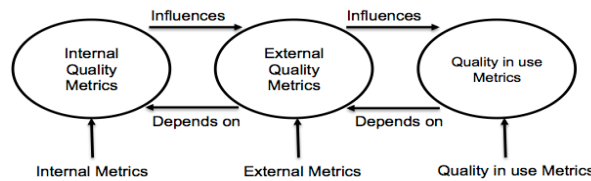

Figure 2. The Relationships between the Quality metrics based on ISO/IEC

To analyze our framework, we consider a case study and the metrics that ISO/IEC provides. The most suitable example is the Busuu project [27]. The Busuu project is an online social network application in which learners can assist each other to improve their language skills. The application provides learning units for twelve different languages, and it can be downloaded for use on mobile phones. This application was designed to enable users to set up a profile and practice (quality metrics of user roles and profiles). Software developers were careful with the learning content that displayed on screens; they targeted as much learning content as possible (quality metric of media type). On the other hand, since the application can be downloaded to mobile phones, users are free to use it wherever network connectivity is available (quality metrics of learning on the move). In addition, each individual user of this application is not only a student of a foreign language, but also a tutor of his or her own mother tongue. One user can communicate and interact with other users (quality metrics of communication, collaboration, and interactivity). However, by using some of the appropriate metrics from ISO/IEC (e.g., functionality, scalability, and service quality), we will be able to measure the quality of m-Learning applications.

Table 1 shows the analysis of the Busuu project based on our framework. In this case, the analysis walks us through from the objectives to the design issues. The purpose of this reverse engineering is to measure the success of the Busuu project and also to determine the design issues our framework can assess and address.

Table 1. Analysis of Busuu Project Based on our Framework.

| Objective | Learning Experience | Learning Contests | Design Issues |
|---|---|---|---|
| **Initial Interaction:**<br><br>1. Exploring, discovering and becoming familiar with the software.<br><br>2. Communicating, interacting, and collaborating with peers by asking and answering questions. | **Rules: Business rules, learning roles:** Different users meet in the context of a simulation.<br><br>**Outcome and Feedback:** Asking questions and getting answers.<br><br>**Goals and objectives:** To become familiar with the application.<br><br>**Conflict, competition, challenge, opposition:** Discussing and challenging opinions (teamwork and new skills).<br><br>**Interaction, blogs, wikis, discussion groups, test, framework:** One to one, one to many, and many to one. | **Identity**: User name and password for each individual user.<br><br>**Learner**: Different users.<br><br>**Activity**: To engage in participatory simulation of a dynamic system.<br><br>**Time-Location**: Co-located same time and different time.<br><br>**Facility**: Different mobile devices.<br><br>**Community**: Different users with different backgrounds using the mobile devices with the support of wireless connectivity can discuss many different topics in order to improve. language skills | **User roles and profiles:**<br><br>**New Users:** Few ideas on how to use the application.<br><br>**Learn on the move:** Mobile devices with the support of wireless connectivity.<br><br>**Media type:** Text, images, comprehensive audio-visual learning material with photos and recordings by native speakers, avoid information overload<br><br>**Communication, collaboration and interactivity:** Users can communicate and collaborate using text, verbal, and video-chat communication support |
| **Learning New Language:**<br><br>By sharing and exchanging information between users provides new knowledge that will help them to improve language skills and to improve the following objectives:<br><br>1. Team building<br>2. Social skills<br>3. New knowledge<br>4. Improved skills. | **Rules: Business rules, learning roles:** Lessons, tutorials, assignments, and group sessions with the support of mobile devices.<br><br>**Goals and objectives:** To get and give answers and to engage in participatory simulation to learn a new language and/or improve language skills.<br><br>**Conflict, competition, challenge, opposition:** Discussing and challenging opinions.<br><br>**Interaction, blogs, wikis, discussion groups, test, framework:** One to one, one to many, and many to one. | **Identity**: Different users.<br><br>**Learner**: Different users.<br><br>**Activity**: Explaining and discussing participative experience.<br><br>**Time-Location**: Co-located same time and different time.<br><br>**Facility**: Different mobile devices.<br><br>**Community**: Different users with different backgrounds using the mobile devices with the support of wireless connectivity can discuss many different topics in order to improve language skills. | **User role:** Participant in group discussion (users).<br><br>**Learn on the move:** Mobile devices with the support of wireless connectivity.<br><br>**Media type:** Text, images, comprehensive audio-visual learning material with photos and recordings by native speakers, vocabulary and key phrases, dialogues, audio, podcasts and PDFs and avoid information overload.<br><br>**Communication, collaboration and interactivity:** Users can communicate and collaborate using text, verbal, and video-chat communication support. |

The results indicate that the Busuu project is successful since it has met all of the requirements of our framework. Also, the success of this project can be determined by the number of downloads. In less than two years, the Bussu project's applications have already been downloaded more than 10 million times [27]. However, one of the goals of our framework is that it should be used to support forward engineering and be used as a design guideline for developing m-Learning applications. Usability has been stated as one of the most important fundamentals of m-Learning applications [28][29]. These weaknesses in an application will prevent users from being efficient, effective, and productive: i) difficult to use, ii) a difficult-to-learn user interface, iii) user interface that is difficult to remember how to reuse; iv) learning content structure that is unclear; v) a process workflow that is difficult to perform [22][30]. The user interface must be effective and easy to use; this will help users to focus on their learning goals, learning content, and activities instead of how the system works; moreover, utilizing design guidelines are vital in developing learning systems [22][31].

### 3.3. Development of Prototype for Assessing the Framework

The interface of the Main Menu of the developed application is shown in Figure 3. The prototype application consists of two user interfaces named Model A and Model B. Model A was designed and developed based on a user interface adopted from the Blackboard website, while Model B was developed following Android SDK recommendations and using our framework as a guideline.

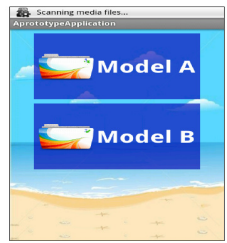

Figure 3. Main Menu of the Prototype Application

The prototype uses familiar terminologies in this prototype application that could be seen on most of WebCT such as "Course Map", "Course Information", "Assignments", "Announcements", "Discussion Board", "Media", "Grades", and "Blogs". This can be seen in Figures 4 and 5 below for Model A (Blackboard based) and Model B (based on our framework) respectively. A comparison of UI for Model A in Figure 4 and UI for Model B in Figure 5 shows that Model B's UI is distinctive and easier to understand with large and self-explaining icons. In contrast Model A's user interface has smaller and less easily understood options. This is also evident in the Blog icons for Model A and Model B in Figure 6 and Figure 7 respectively: Model B is distinctive and easier to understand. The purpose for developing these different user interfaces is to find the best way to design and develop a user-friendly user interface for mobile applications to increase the usability level of these applications.

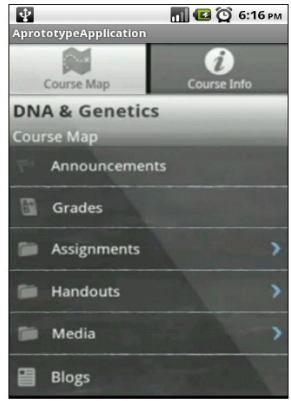

Figure 4. Model A based on Blackboard

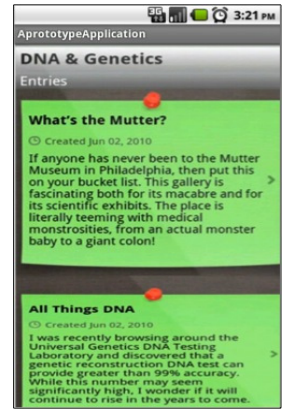

Figure 6. Blog's Icon in Model A

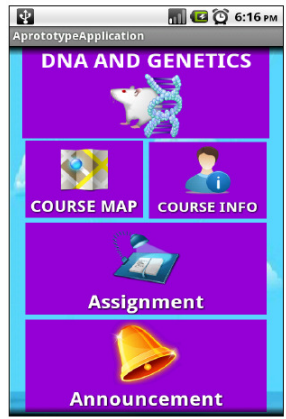

Figure 5. Model B based on Android
Recommendations and our framework as a
guideline

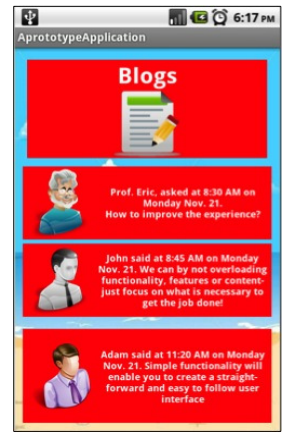

Figure 7. Blog's Icon in Model B

## 4.RESEARCH  METHODOLOGY

### 4.1Research participants

The sample population of the result was undergraduate students at Western University in Canada. The population was limited to second- and third-year students in the undergraduate program in the Software Engineering department. Questionnaires were handed to the students and a total of 96 completed questionnaires from the complete population of the present study. These students were in the 19-23 age group and were both male and female students enrolled in the software engineering program
.

## 4.2. Instrument

The instrument used for analysis was a questionnaire survey handed out to interested participants who took part in a live assessment of the two Models of m-Learning developed as a part of this study. The m-Learning platform prototype developed for this study has used for a heuristic evaluation as a technique to measure usability factors. Heuristic evaluation is an engineering method for easy, quick, and cheap evaluation of a user interface design [11]. It is known as one of the most popular usability inspection methods, and it is done as a systematic inspection of user interface design for usability [11]. As mentioned, a usability questionnaire was conducted to evaluate the usability issues of the prototype application among 96 students. However, using this technique and giving the participants real mobile devices, allowed participants to use this application to share opinions regarding their experiences while interacting with the real prototype application. Participants rated each question from 1 to 5 on a Likert scale (1=very easy, 5=very difficult).

After they were given the questionnaire survey form they were provided with real smartphones. The main goal here was to compare the two user interfaces in the four sections and determine which one of these interfaces is better, in terms of usability sub-characteristics.

## 4.3. Assessment Method

Upon collecting the data, the level of usability was investigated by evaluating the application user interfaces, which include ease of use, user satisfaction, and attractiveness and learnability. A comparison will be made to determine the more user-friendly interface between the two models that form a part of the prototype.

# 5. DATA ANALYSIS

The data analysis aims at comparing user interfaces Model A and Model B for smartphone applications. The specific areas of comparing the two models will be: ease of use, user satisfaction, attractiveness, and learnability. The sample contains 96 students who used the UI for the application and finished the survey correspondingly. First a descriptive analysis of the data collected is presented. After this a comparative analysis of the responses of participants using Model A and Model B will be presented based on the four aspects mentioned earlier. This will be followed by the reliability and validity of data. Finally, the association analysis is presented to ensure that the user evaluates both models independently.

## 5.1. Descriptive Analysis of Data

For the purpose of analysis, the participant responses for both user interfaces (Model A and Model B) have been gathered on four aspects of usage: ease of use, user satisfaction, attractiveness, and learnability. Table 2 below presents the mean and standard deviations of the participant responses for the four aspects of usage.

Table 2. Statistical Analysis of Data

| Mean and SD | Usability sub-characteristics | | | |
|---|---|---|---|---|
| | Ease of Use | User satisfaction | Attractiveness | Learnability |
| Model A Mean | 2.47 | 2.23 | 3.16 | 2.17 |
| Model A SD | 0.37 | 0.27 | 0.27 | 0.1 |
| Model B Mean | 1.74 | 1.76 | 2.05 | 1.71 |
| Model B SD | 0.14 | 0.19 | 0.13 | 0.09 |

## 5.2. Comparative Analysis of Data

As multiple questions were used to evaluate both models in each section the same weight was given to every question for overall evaluation.

The statistics of "ease of use" sub-characteristic from Table 2 shows that Model B, with a mean of 1.74, has a lower score than Model A, with a mean of 2.47. This means that participants consider the user interface of Model B to be easier to use as compared to the user interface of Model A. The results are corroborated by the standard deviation (SD) of the two models. Model B with SD 0.14 is lower than the SD of Model A, which is 0.37. This shows that the Model B responses are more consistent.

The statistics of "user satisfaction" sub-characteristic from Table 2 shows that Model B, with a mean of 1.76, has a lower score than Model A, with a mean of 2.23. This means that participants are more satisfied with the user interface of Model B as compared to the user interface of Model A. The results are corroborated by the standard deviation of the two models. Model B with SD 0.19 is lower than the SD of Model A, which is 0.27. This shows that the Model B responses are more consistent.

The statistics of "attractiveness" sub-characteristic from Table 2 shows that Model B, with a mean of 2.05, has a lower score than Model A, with a mean of 3.16. This means that participants consider the user interface of Model B to be better looking as compared to the user interface of Model A. The results are corroborated by the standard deviation of the two models. Model B with SD 0.13 is lower than the SD of Model A, which is 0.27. This shows that the Model B responses are more consistent.

The statistics of "learnability" sub-characteristic from Table 2 shows that Model B, with a mean of 1.71, has a lower score than Model A, with a mean of 2.17. This means that participants consider that it is easier to learn a topic utilizing the user interface of Model B as compared to learning with the user interface of Model A. The results are corroborated by the SD of the two models. Model B with SD 0.09 is lower than SD of Model A, which is 0.1. This shows that the Model B responses are more consistent.

The results of the four sub-characteristics can be averaged to find the overall comparative performance of Model A and Model B.

## 5.3. Reliability and validity of data

In order to validate our intuitive understanding regarding the comparison between Model A and Model B, a paired T-test and F-test is used to test the different scores between Model A and Model B, as well as the variance differences between the two models. The hypothesis for the test shown in Table 3 below is that Model B performs the same as Model A. This is tested for each of the sub-characteristics, resulting in a total of four hypotheses (H1, H2, H3, and H4) to be tested. In addition to the four hypotheses, a general case hypothesis, H5, is also created by averaging the results from the four sub-characteristics.

Table 3. Hypothesis for usability sub-characteristics.

| Sub Characteristics | Hypothesis Statement to be Tested |
|---|---|
| Ease of Use | |
| User satisfaction | Model B performs the same as Model A |
| Attractiveness | |
| Learnability | |

The outcomes of the reliability and validity analysis are shown in Table 4. For each of the usability sub-characteristics the test statistics and confidence intervals for both T-test and F-test are given in the table.

Table 4. Analysis of data using T-test and F-test.

| Hypotheses | Paired T-Test | Paired F-Test |
|---|---|---|
| | Test Statistics | Test Statistics |
| Ease of Use | 15.0* | 1.32* |
| User satisfaction | 10.5* | 1.30* |
| Attractiveness | 20.7* | 0.70* |
| Learnability | 7.9* | 1.37* |
| General Case | 27.3* | 1.32* |

(*) Significant at p-value < 0.05, and (**) Insignificant at p-value > 0.05

• <u>Ease of Use</u> – From Table 4 above it can be seen that this sub-characteristic, the hypothesis that Model A and Model B perform the same, is rejected. The hypothesis is rejected because of the test statistic that is statistically significant at p-vale < 0.05. This indicates that Model A and Model B have significantly different performance levels from this perspective. In order to perform variance and consistency analysis, again the hypothesis that Model A and Model B perform the same is used. In this case, too, the hypothesis is rejected as the test statistic is significant at p < 0.05. This means that the consistency situation for Model A and Model B is different in terms of ease of use.

• <u>User Satisfaction</u> – From Table 4 above it can be seen that this sub-characteristic, the hypothesis that Model A and Model B perform the same, is rejected. The hypothesis is rejected because of the test statistic that is statistically significant at p-vale < 0.05. This indicates that Model A and Model B have significantly different performance levels from the user satisfaction perspective. In order to perform variance and consistency analysis, again the hypothesis that Model A and Model B perform the same is used. In this case, too, the hypothesis is rejected as the test statistic is significant at p < 0.05. This means that the consistency situation for Model A and Model B is different in terms of user satisfaction.

• <u>Attractiveness</u> – From Table 4 above it can be seen that this sub-characteristic, the hypothesis that Model A and Model B perform the same, is rejected. The hypothesis is rejected because of the test statistic that is statistically significant at p-vale < 0.05. This indicates that Model A and Model B have significantly different performance levels from the attractiveness perspective. In order to perform variance and consistency analysis, again the hypothesis that Model A and Model B perform the same is used. In this case, too, the hypothesis is rejected as the test statistic is significant at p < 0.05. This indicates that Model A shows more consistency than Model A in terms of responses for attractiveness. When the reasons for this result are analyzed, it is found that attractiveness is more of a subjective judgement and is dependent on an idiosyncratic factor for each user. Yet another reason for this outcome can be due to the fact that the evaluation of Model A was more concentrated and focused. Several of the questions in this section were regarding the colour of the interface. Most of the user interface of Model B has different shades of red, and the feedback from the users showed an aversion to the colour with several participants suggesting this colour should not be used.

• <u>Learnability</u> – From Table 4 above it can be seen that for this sub-characteristic, the hypothesis that Model A and Model B perform the same is rejected. The hypothesis is rejected because of the test statistic that is statistically significant at p-vale < 0.05. This indicates that Model A and Model B have significantly different performance levels from the learnability perspective. Thus these statistics validate our intuitive understanding that Model B performs better than Model A (as per the comparative analysis of data in the previous section). In order to perform variance and consistency analysis, again the hypothesis that Model A and Model B perform the same is used. In this case, too, the hypothesis is rejected as the test statistic is significant at $p < 0.05$.

• General Validation - From Table 4 above it can be seen that from the overall general perspective, the hypothesis that Model A and Model B perform the same is rejected. The hypothesis is rejected because of the test statistic that is statistically significant at p-vale < 0.05. This indicates that Model A and Model B have significantly different performance levels from an overall general perspective. In order to perform variance and consistency analysis, again the hypothesis that Model A and Model B perform the same is used. In this case, too, the hypothesis is rejected as the test statistic is significant at $p < 0.05$. This means that the consistency situation for Model A and Model B is different in terms of user satisfaction. This indicates that Model B shows more consistency than Model A

## 5.4. Association Analysis

For the purpose of this study, the evaluation of both Models A and B has been conducted simultaneously. Hence, it is essential that the association levels between the evaluation of Models A and B are tested. This is to check whether the user evaluates Model A and Model B independently or not. The test is performed on the five hypotheses (H1, H2, H3, H4, and H5) for each of the four usability sub-characteristics, as depicted in Table 3. One of the methods to check for association levels in the parametric statistics is to find the Pearson correlation coefficient between Model A and Model B. Since this is a statistical test, the p-value – i.e., the probability of obtaining a test-statistic – is observed. The lower the p-value, the less likely that the result if the null hypothesis is true and, accordingly, the more "significant" is the result in terms of its statistical significance [36]. The Spearman Coefficient is the counterpart of the Pearson Coefficient in non-parametric analysis, which defines the correlation between the two models. The results for the Pearson correlation coefficient and the Spearman Correlation Coefficient are displayed in Table 5 below.

Table 5. Analysis of the Data Using Pearson and Spearman Correlation Methods.

| Hypotheses | | Correlation Coefficients | |
|---|---|---|---|
| | Section | Pearson Correlation Coefficient | Spearman Correlation Coefficient |
| H1 | Ease of Use | 0.06** | 0.09* |
| H2 | User satisfaction | 0.20* | 0.21* |
| H3 | Attractiveness | -0.010** | -0.008** |
| H4 | Learnability | 0.22* | 0.24* |
| H5 | General Case | 0.16* | 0.17* |

(*) Significant at p-value < 0.05, and (**) Insignificant at p-value > 0.05

• Ease of Use – From Table 5 above it can be seen that the outcomes of the analysis for this sub-characteristic regarding the association between Model A and Model B are not consistent. In fact the outcomes are contradictory. The Pearson Correlation Coefficient test is positive (0.06) at P-Value > 0.05, hence, the hypothesis cannot be rejected. The Spearman Correlation Coefficient test, on the other hand, is (0.09) at P-Value < 0.05; therefore, the hypothesis is rejected. However, if we take a closer look at the P-value, we can see that the P-Value for the Spearman coefficient is (0.027), which is close to the boundary of 0.05. This situation indicates the inaccuracy of this test. Another possible cause for this situation is that the Spearman coefficient is a non-parametric association estimation. Therefore, the sample size plays a very important role in the accuracy of this coefficient estimation. Our sample is limited to the 96 students who participated in the survey. The limited sample size will make the coefficient less accurate. Thus in this case, the outcome of the Pearson coefficient, which indicate that the evaluation of these two models is independent, must be accepted as the actual outcome.

• <u>Usage Satisfaction</u> – In the user satisfaction section, The Pearson Correlation Coefficient between Model A and Model B is positive (0.20) (P-Value < 0.05), and the Spearman Correlation Coefficient test is also positive (0.21) (P-Value < 0.05). Since the P-Values for the two Coefficients indicate the significance of the test, it can be seen that both the Pearson Coefficient and the Spearman Coefficient indicate that we should reject the hypothesis. Therefore, the evaluation of Model A has a positive relationship with that of Model B, which indicates that the users who evaluate Model A with a higher score will also tend to evaluate Model B with a higher score.

• <u>Attractiveness</u> – From the outcomes of the attractiveness section, we find that the P-value for the Pearson Coefficient and the Spearman Coefficients are negative (-0.010, -0.008) with P-Value > 0.05. Since the P-Values for the two Coefficients indicate the insignificance of the test, both the Pearson Coefficient and Spearman Coefficient indicate that the null hypothesis about the independence situation between these two Models cannot be rejected. However, since attractiveness is purely a subjective judgment towards the models, the evaluation between Model A and Model B will be independent.

• <u>Learnability</u> – For the learnability, the Pearson correlation coefficient between Model A and Model B is positive (0.22) (P-Value < 0.05). Furthermore, the Spearman Correlation Coefficient test is also positive (0.24) (P-Value < 0.05); the p-values for the two coefficients indicate the significance of the test. Therefore, the Pearson Coefficient and Spearman Coefficient suggest that we should reject the hypothesis. Moreover, the evaluation of Model A has a positive relationship with that of Model B, which suggests that users who evaluated Model A with a higher score also tend to evaluate Model B with a higher score.

• <u>General Case</u> – Finally, in the general case section, the Pearson Correlation Coefficient between Model A and Model B is positive (0.16) (P-Value < 0.05), and the Spearman Correlation Coefficient test is also positive (0.17) (P-Value < 0.05). Since the P-Values for the two coefficients indicate the significance of the test, the Pearson Coefficient and Spearman Coefficient indicate that we should reject our hypothesis. Also, the evaluation for Model A has a positive relationship with that of Model B, suggesting that the users who evaluate Model A with a higher score also tend to evaluate Model B with a higher score.

## 6.DISCUSSION OF RESULTS

The prime objective of this work is to propose a conceptual framework for measuring the usability aspects of m-Learning. Further, the framework was used as a guideline to build a prototype model of a prototype application for smartphones. This model was compared with another model based on the Blackboard website on the basis of questionnaire responses from a survey conducted at the University of Western Ontario. The comparative analysis shows that the distribution of Model B was smaller than that of Model A. According to our pre-defined options for the questions, practitioners tend to rank Model B better than model A. In order to support our intuitive idea of the data, a validation test of the Models was performed.

As the evaluation of the two Models A and B was paired together, the paired T-test was used to evaluate the difference between the mean of Model A and Model B for each question and the general case. The F-test was then used to test whether the variance of Model A will be the same as that of Model B in different sections. The results of the analysis showed that the p-value of the T-test and F-test was less the 0.05, the null hypothesis (i.e., the difference between the mean of Models A and B is 0) should be rejected.

The samples were further analyzed using the association test to examine whether there is relationship between when students evaluate Model A and Model B. The key statistics used for determining this were the Pearson Correlation Coefficient and the Spearman Correlation Coefficient. After testing the hypothesis it was found that for some questions, students evaluate Model A and Model B independently. However, the overall results demonstrated the existence of a positive correlation between the evaluation of Model A and Model B. This means that students who evaluate Model A higher tend to evaluate Model B higher.

In conclusion, the analysis pointed clearly that Model B would give a superior usability performance as compared to Model A in terms of the four sub-characteristics used to evaluate usability: ease of use, user satisfaction, attractiveness, and learnability.

# 7. CONCLUSION

M-learning applications must be easy to use, learnable, understandable, and attractive as well as provide an enjoyable experience for users. It is important to meet usability needs for the m-Learning applications, since user interface plays the most important role for each individual interaction between the user and his/her smartphone application. However, little attention has been paid to researching the usability assessment in the area of m-Learning application design.

One of the key contributions of this work is the development of a conceptual framework for measuring the quality aspects of m-Learning. In addition, a prototype application for smartphones using the Java Language and an Android Software Development Kit has also been developed by following the proposed framework as a guideline. We also conducted a questionnaire survey at Western University to compare this prototype with a model based on the Blackboard website. While performing the comparative analysis, it was discovered that our model – developed using our framework as a guideline – performed better than the model developed based on the Blackboard website in terms ease of use, user satisfaction, attractiveness, and learnability.

This framework can be used as a guideline to support forward and reverse engineering and for use in the future while developing mobile applications. The key limitation of the analysis is the limited demographics used (comprising only undergraduate software engineering students) while conducting the research questionnaire, which means that the results cannot be deemed as universal. Future research could involve the quality perspective of the learning material itself. The research can also be extended to assess the design issues of learning-on-the-move, collaboration, and communication and interactivity.

## Authors

Mr. Abdalha Ali is graduated with a Masters degree in Software Engineering in the Department of Electrical and Computer Engineering at Western University. He received his B.A.Sc. in Electrical and Computer Engineering from the University of Al Jabal Al Gharbi, Libya, in 2005. His research interest includes technology tools for teaching and the use of mobile devices in education. Mr. Ali worked for more than 3 years in the Higher Ministry of Education in Libya, where he gained good experience in project Management. He can be reached at: aali252@uwo.ca

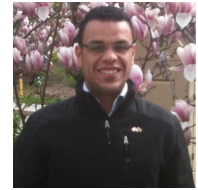

Mr. Muasaad Alrasheedi is a Ph.D. Candidate in the department of Electrical and Computer Engineering (Software Engineering Program) at Western University, London, Canada. Mr. Alrasheedi has a Bachelor of Science degree in Information Technology and Computing From Arab Open University, Saudi Arabia and Master of Engineering in Technology Innovation Management from Carleton University, Ottawa, Canada. His research interest is in the mobile learning and emerging educational technology. He can be reached at: malrash@uwo.ca.

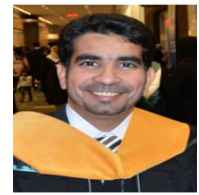

Dr. Abdelkader Ouda is an Assistant Professor in the Department of Electrical and Computer Engineering at Western University. He received his B.Sc. (Hons) and M.Sc. degrees in Computer Science and Pure Mathematics from Ain Shams University, Cairo, Egypt, in 1986, and 1993 respectively, and his Ph.D. degree in Computer Science from the Western University, Canada in 2005. Dr. Ouda has been studying, researching, and teaching software engineering at Western University for 12 years. His current research is mainly in Information and networking security. Dr. Ouda was working for over 14 years in the software development industry, involving project management, system analysis/consulting, and database design. He can be reached at: aouda@uwo.ca.

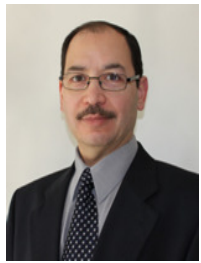

Dr. Luiz Fernando Capretz is a Professor of Software Engineering and Assistant Dean (IT & e-Learning) at Western University, London, Canada. His research interests include software engineering, technology-enhanced learning, human factors in software engineering, and software engineering education. Dr. Capretz has a PhD in computing science from the University of Newcastle upon Tyne. He is a senior member of the IEEE, a distinguished member of the ACM, an MBTI certified practitioner, and a Professional Engineer in Ontario (Canada). He can be reached at: http://www.eng.uwo.ca/people/lcapretz/default.htm. Email: lcapretz@uwo.ca

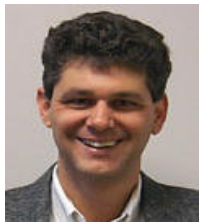